\documentclass[%
 reprint,superscriptaddress,
 amsmath,amssymb, nofootinbib,
 aps,twocolumn,pra
]{revtex4-1}
\usepackage{graphicx}% Include figure files
\usepackage{appendix}
\usepackage{multirow}% For tables
\usepackage{amsfonts}
\usepackage{bbold} % For identity operator
\usepackage{verbatim} % typewriter style for code
\usepackage{ulem} % cross out text
\usepackage[dvipsnames]{xcolor}
\usepackage{xifthen}

\newcommand{\ket}[1]{| #1 \rangle}

\newcommand{\arxiv}[2][]{\ifthenelse{\isempty{#1}}{\href{http://arxiv.org/abs/#2}{{\tt arXiv:\allowbreak{}#2}}} {\href{http://arxiv.org/abs/#2}{{\tt arXiv:\allowbreak{}#2 [#1]}}}}

\begin{document}

\title{Quantum Random Number Generator with Internal Consistency Check and Public Verification}

\author{Rodrigo Piera\smallskip}
\affiliation{Quantum Research Centre, Technology Innovation Institute, Abu Dhabi, United Arab Emirates}
\author{Gianluca De Santis \smallskip}
\affiliation{Quantum Research Centre, Technology Innovation Institute, Abu Dhabi, United Arab Emirates}
\author{Agustin Sanchez \smallskip}
\affiliation{Quantum Research Centre, Technology Innovation Institute, Abu Dhabi, United Arab Emirates}
\author{Yury Kurochkin\smallskip}
\affiliation{Quantum Research Centre, Technology Innovation Institute, Abu Dhabi, United Arab Emirates}
\author{James A. Grieve\smallskip}
\affiliation{Quantum Research Centre, Technology Innovation Institute, Abu Dhabi, United Arab Emirates}

%\altaffiliation[Also at ]{Physics Department, XYZ University.}

%\date{9 December 2020}

% \begin{description}
% \item[PACS numbers]

% \pacs{Valid PACS appear here}
% \end{description}

\begin{abstract}
Quantum Random Number Generators provide true physical randomness based on quantum processes, essential for cryptographic and scientific applications. However, practical implementations face challenges in robustness and verifiability: ensuring that the entropy source remains secure and stable over time, and enabling independent confirmation of randomness quality without compromising security.
We present a system based on a simple looped beam splitter architecture that uses only passive optical components. The device features an intrinsic self-testing mechanism derived from the stability of detection-probability ratios, allowing continuous validation of correct operation.
In addition, the same physical process generates two independent random sequences with identical entropy: a private sequence, used for secure applications, and a public one, enabling external statistical verification with zero mutual information between them.
This approach demonstrates that robust, self-testing, and publicly verifiable quantum randomness can be achieved with minimal optical complexity without jeopardizing security.
\end{abstract}

\maketitle

\section{Introduction}

Reliable randomness underpins modern secure communications and computational protocols. Quantum mechanics provides a fundamentally non-deterministic source of randomness, making it ideal for applications where unpredictability and integrity are critical \cite{calude2008quantum,khrennikov2016probability}. Quantum Random Number Generators (QRNGs) exploit such intrinsic quantum processes to produce unpredictable outputs for use in cryptography \cite{hayes2001computing,schindler2009random}, computer simulations \cite{metropolis1949monte}, and random sampling tasks \cite{konstantinou2004electronic}. However, practical implementations of QRNGs rely on physical components that can deviate from their idealized theoretical models, potentially introducing biases or correlations that degrade the quality of the randomness \cite{nie2016experimental,NIST80090B}. In both classical and quantum devices, the complexity of the design amplifies these risks by adding hidden dependencies and points of failure. Simplified architectures mitigate these vulnerabilities by reducing the number of active components and operational parameters, thereby enhancing reliability, stability, and verifiability.

Even in simplified designs, hardware degradation or misalignment can still compromise the quality of the generated randomness over time \cite{fischer2012closer}. In practice, any drift in component performance, such as detector efficiency or optical coupling, can distort the expected quantum statistics. For this reason, recent standards such as AIS 31 \cite{AIS31} and NIST SP 800-90B \cite{NIST80090B} emphasize the inclusion of continuous health-testing mechanisms capable of detecting anomalies in real time. Quantum randomness originates from processes governed by well-defined probability distributions \cite{mandel1995optical,khrennikov2016probability}, therefore QRNGs can be continuously monitored by verifying that the measured statistics remain consistent with the predicted quantum model. This capability forms the basis of an internal consistency check, ensuring that the randomness source operates within its expected quantum regime and that any deviation can be promptly detected.

In addition, public verifiability, that is, the ability of parties to confirm the quality of the generated randomness without revealing the private random sequence, has emerged as a desirable property for QRNGs \cite{jacak2020quantum,islam2024privacy,piera2024delegated}. This functionality is especially relevant for applications that require externally trusted randomness, such as certified public randomness beacons, blockchain protocols, or remote cryptographic key generation, where independent auditing is essential. In our approach, we generate an auxiliary public sequence that can be openly tested to verify the performance of the source, while the private sequence remains secure. 

%Although both sequences arise from mutually exclusive detection events and therefore share the same underlying probabilities, any potential information leakage is fully captured by the entropy bound used in post-processing.

Quantum random number generation can be broadly classified into three categories \cite{ma2016quantum}: device-independent schemes based on entanglement tests \cite{colbeck2011private, fehr2013security, liu2018high}, semi-device-independent protocols based on physical constraints such as energy or dimensionality \cite{avesani2021semi, cheng2024semi,an2018experimental}, and trusted-device implementations \cite{sanguinetti2014quantum,nie2016experimental}. Our work belongs to the latter class, but introduces two complementary layers of verification: an internal statistical consistency check and an external public validation channel that bridges the gap between fully trusted and externally auditable QRNG architectures.

The system operates using only passive optical components, and its detection-time statistics provide a direct signature of correct functioning. By partitioning detection events into two statistically identical but independent sequences, we demonstrate the generation of private and public random streams with identical entropy. This architecture enables robust, secure, and externally verifiable quantum randomness without the need for complex feedback or active stabilization.

\section{Background}

\begin{figure}[ht]
    \centering
    \includegraphics[width=8.4cm]{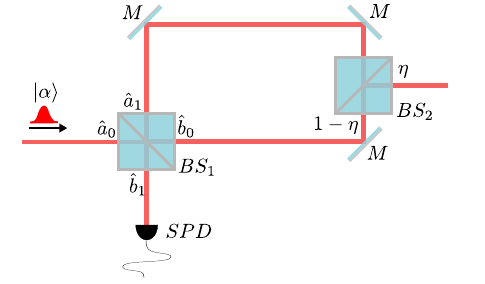}
    \caption{Scheme of the loop-based QRNG. BS: Beam Splitter; M: Mirror; SPD: Single Photon Detector; \(\eta\): effective system loss.}
    \label{fig:setup_1}
\end{figure}

The proposed QRNG, illustrated in Fig.~\ref{fig:setup_1}, relies on time-resolved detections from an optical loop. Photons are injected into a beam splitter ($BS_1$) with reflectivity \(r\) and transmissivity \(t = 1 - r\). One output port, \(b_1\), is directed to a single-photon detector (SPD), while the other output, \(b_0\), is fed back to the opposite input arm \(a_1\), thus creating a loop. Losses in the system are modeled by an additional fictitious beam splitter ($BS_2$) with reflectivity \(\eta\).

Each optical pulse incident on $BS_1$ can either be reflected toward the SPD and detected directly, or transmitted into the loop, and re-enter $BS_1$ from the opposite input arm after one round trip. 
% Because the light re-enters from the other input port, the \emph{roles of reflection and transmission swap} on successive passes: to exit toward the same detector port \(b_1\), the first pass requires reflection, while subsequent passes require transmission. This symmetry gives rise to two physically distinct cases—direct and looped detections—governed by different expressions for the field intensity reaching the detector.

For a weak coherent pulse \( \ket{\alpha} \), the probability of at least one detection event after the \(l\)-th loop follows the exponential form:
\begin{equation}
    p^{l}_{\text{click}} = 1 - e^{-\beta_l},
\end{equation}
where the mean photon number at the detector, \(\beta_l\), depends on the optical path taken:
\begin{equation}
    \beta_l =
    \begin{cases}
        |\alpha|^2 r, & l = 0, \\[4pt]
        |\alpha|^2\, t^2\, r^{\,l-1}\,(1 - \eta)^{l}, & l \ge 1.
    \end{cases}
    \label{eq:beta_definition}
\end{equation}
The term \(\beta_{l=0}\) corresponds to photons that are reflected directly toward the detector without entering the loop. Their detection probability depends only on the beam-splitter reflectivity \(r\).  
In contrast, photons that complete one or more round trips (\(\beta_{l \ge 1}\)) experience both reflections and transmissions at the beam splitter, as well as cumulative propagation losses. The factor \(t^2 r^{\,l-1}\) accounts for the beam splitter interactions, and \((1-\eta)^l\) captures the attenuation accumulated after \(l\) circulations.

% Although the functional form of Eq.~\eqref{eq:beta_definition} is valid for all \(l\), the \(l=0\) contribution does not belong to the same sequence as the looped events. It lacks the reflection–transmission alternation and loss dependence that define the geometric scaling of \(\beta_l\) for \(l\ge1\). Consequently, including \(l=0\) detections would break the constant-ratio relation among looped probabilities that underlies the internal consistency check discussed below. For this reason, only events with \(l \ge 1\) are used for the self-verification and bit generation.

The geometric scaling for the terms \(\beta_{l \ge 1}\) is at the foundation of the constant-ratio relation used for the internal consistency check discussed below. For this reason, events where \(l=0\) are discarded for the self-verification and bit generation stages.

To ensure single-event timing, we post-select records with exactly one detection and label each event with its corresponding loop index \(l\) (arrival time). The probability of detecting only one event at loop \(l\) is:
\begin{equation}
    P_{l} = p^{l}_{\text{click}} \prod_{i \neq l} \bigl(1 - p^{i}_{\text{click}}\bigr).
\end{equation}

% i.e., a detection at \(l\) and no detections at the other loop windows considered.

In the weak-coherent regime (\(\beta_l \ll 1\)), the ratio of consecutive single-detection probabilities simplifies to:
\begin{equation}
    \frac{P_{l+1}}{P_{l}} =  \frac{(1 - e^{-\beta_{l+1}}) e^{-\beta_{l}}}{(1 - e^{-\beta_{l}}) e^{-\beta_{l+1}}}
    \approx \frac{\beta_{l+1}}{\beta_{l}} = r(1 - \eta).
    \label{eq:ratio_sigle_det}
\end{equation}
Under stable operation, this ratio remains constant, thereby enabling an internal consistency check. By monitoring \(P_{l+1}/P_l\) over time, the device statistically validates correct functioning without auxiliary calibration.

We partition the time-resolved outcomes into two disjoint subsets to form two binary sequences: \(S_{Q_1}\) (private) from \(l \in [1,2]\), and \(S_{Q_2}\) (public) from \(l \in [3,4]\). We assign the bit “0” to \(l = \{1,3\}\), and the bit “1” to \(l = \{2,4\}\). Both subsets arise from the same physical process, and the ratio in Eq.~\eqref{eq:ratio_sigle_det} is constant; hence, the two subsets have identical entropy parameters:
\begin{equation}
    \frac{P_1}{P_{1}+P_{2}}  \approx \frac{1}{1 + r(1 - \eta)} \approx \frac{P_{3}}{P_{3}+P_{4}}.
\end{equation}
The public sequence \(S_{Q_2}\) is used for external statistical verification, while the private sequence \(S_{Q_1}\) remains secret. This choice is made because events that generate \(S_{Q_1}\) are more probable than the ones for \(S_{Q_2}\). Therefore, for a given acquisition time, the private sequence will be bigger than the public one. Any dependence between them is already accounted for via the min-entropy bound used in post-processing.

For the private sequence \(S_{Q_1}\), the per-event min-entropy is:
\begin{equation}
    H_{\text{min}}(S_{Q_1}) = \log_2\!\left(\frac{P_{1}+P_{2}}{P_{1}}\right),
\end{equation}
while the probability that a pulse yields a usable private event is \(P_{\mathrm{tot}}^{(Q_1)} = P_1 + P_2\). The public sequence has the same functional form.

Our operational goal is to maximize the expected number of extractable random bits $b$ as a function of the reflectivity of $BS_1$, for given values of $\alpha$ and $\eta$.
The figure of merit we optimize is:

\begin{equation}
    b := P_{\mathrm{tot}}^{(Q_1)} \cdot H_{\text{min}}(S_{Q_1})
    = (P_1+P_2)\,\log_2\!\left(\frac{P_{1}+P_{2}}{P_{1}}\right).
    \label{eq:bits_per_pulse_metric}
\end{equation}

This expression highlights the trade-off introduced by the beam-splitter reflectivity \(r\). Increasing \(r\) tends to equalize \(P_1\) and \(P_2\), thereby increasing the min-entropy per event, but at the same time directs more photons toward the immediate reflection output (\(l=0\)), reducing the number of pulses that enter the loop and contribute to \(P_1 + P_2\). Conversely, smaller values of \(r\) favor a higher loop rate but yield a more biased distribution and therefore a lower entropy per bit. 

\begin{figure}[ht]
    \centering
    \includegraphics[width=8.3cm]{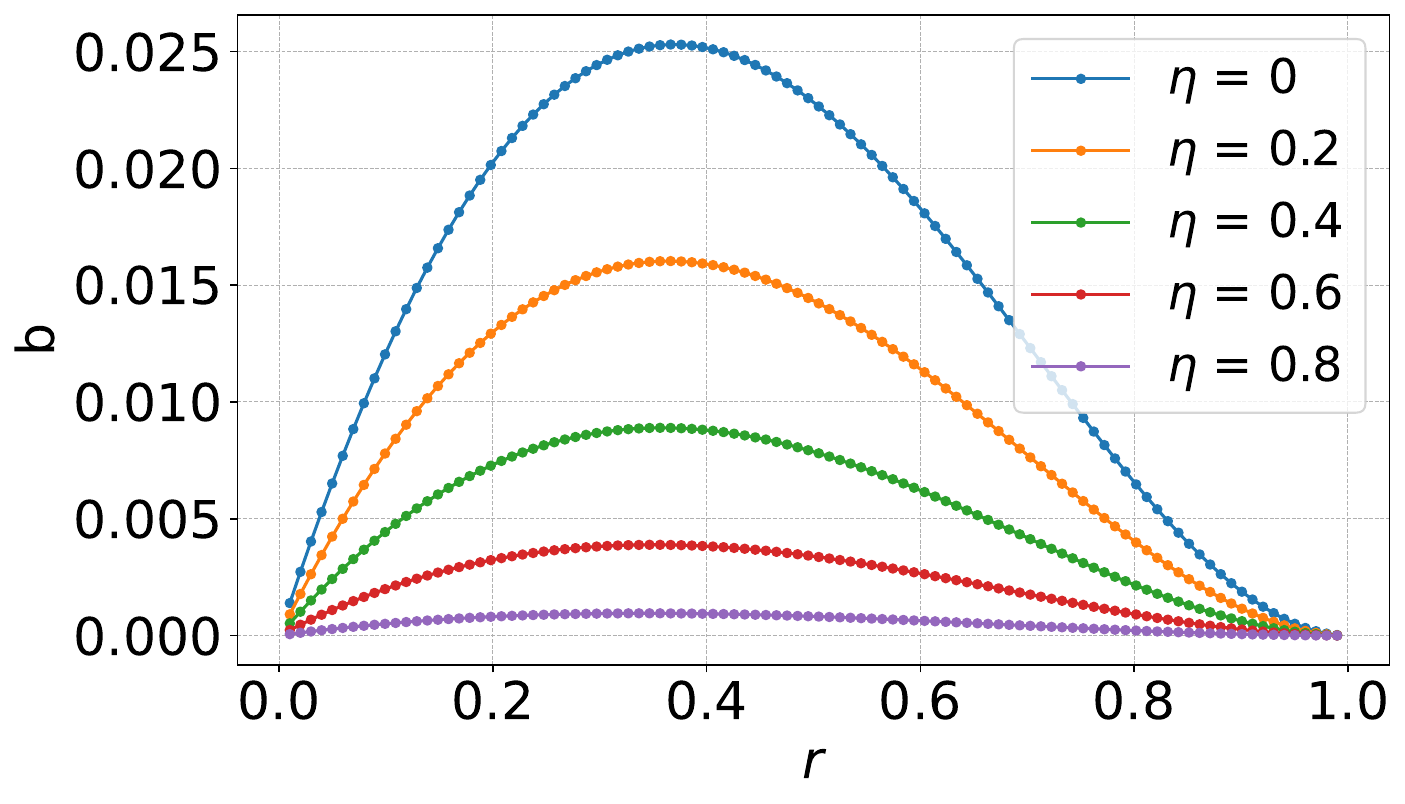}
    \caption{Expected extractable bits per pulse \(b(r;\eta,|\alpha|^2)\) versus reflectivity \(r\) for several loss values \(\eta\). Curves are obtained from \((P_1+P_2)\log_2\!\bigl((P_1+P_2)/P_1\bigr)\) with \(P_l\) defined in the main text.}
    \label{fig:plot}
\end{figure}

Fig.~\ref{fig:plot} illustrates the simulated extractable bits per pulse as a function of the beam splitter reflectivity \(r\) for several system loss values \(\eta\), calculated using Eq.~\eqref{eq:bits_per_pulse_metric}. Higher reflectivity increases the entropy, but it simultaneously decreases the overall usable rate. The extractable bit rate per pulse is thus not maximized at either extreme of \(r\), but rather lies in an intermediate region where these opposing effects balance.

%For typical parameters, a near-balanced beam splitter (\(r \approx 0.5\)) provides close-to-optimal performance.

\section{Experiment and Results}

\begin{figure}[ht]
    \centering
    \includegraphics[width=8.4cm]{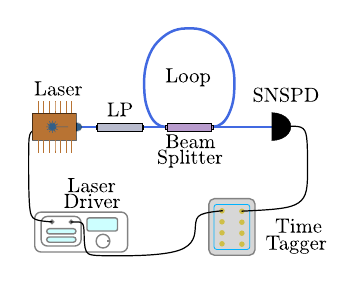}
    \caption{Experimental setup of the QRNG device. A pulsed 1310~nm laser is attenuated and injected through a linear polarizer (LP) into a $40/60$ fiber beam splitter. The reflected output ($b_1$) is sent to a single-photon detector (SNSPD), while the transmitted output ($b_0$) is fed back to the other input port ($a_1$), forming a loop.}
    \label{fig:experimental_setup}
\end{figure}

The experimental setup, depicted in Fig.~\ref{fig:experimental_setup}, employs a pulsed laser at $1310 \, \text{nm}$ operating at a repetition rate of $3 \, \text{MHz}$. The output is attenuated to an average photon number per pulse of $\alpha = 0.33$, ensuring operation in the weak-coherent regime. The pulsed light passes through an in-fiber linear polarizer (LP) and is injected into the input port $a_0$ of a $40/60$ fiber beam splitter. Output $b_1$ of the beam splitter is connected to a superconducting nanowire single-photon detector (SNSPD), while output $b_0$ is connected to the input port $a_1$, forming the feedback loop.

The loop length was adjusted to produce a round-trip time of $33 \, \text{ns}$, longer than the detector's dead time of $25 \, \text{ns}$, ensuring that consecutive loop detections could be distinguished. We post-selected events with a single detection click within the temporal window corresponding to \(l \in [1,4]\) loop transits. A Swabian Time Tagger synchronized the laser trigger and detection signals, enabling precise time stamping of each photon arrival.

\begin{figure}[ht]
    \centering
    \includegraphics[width=8.3cm]{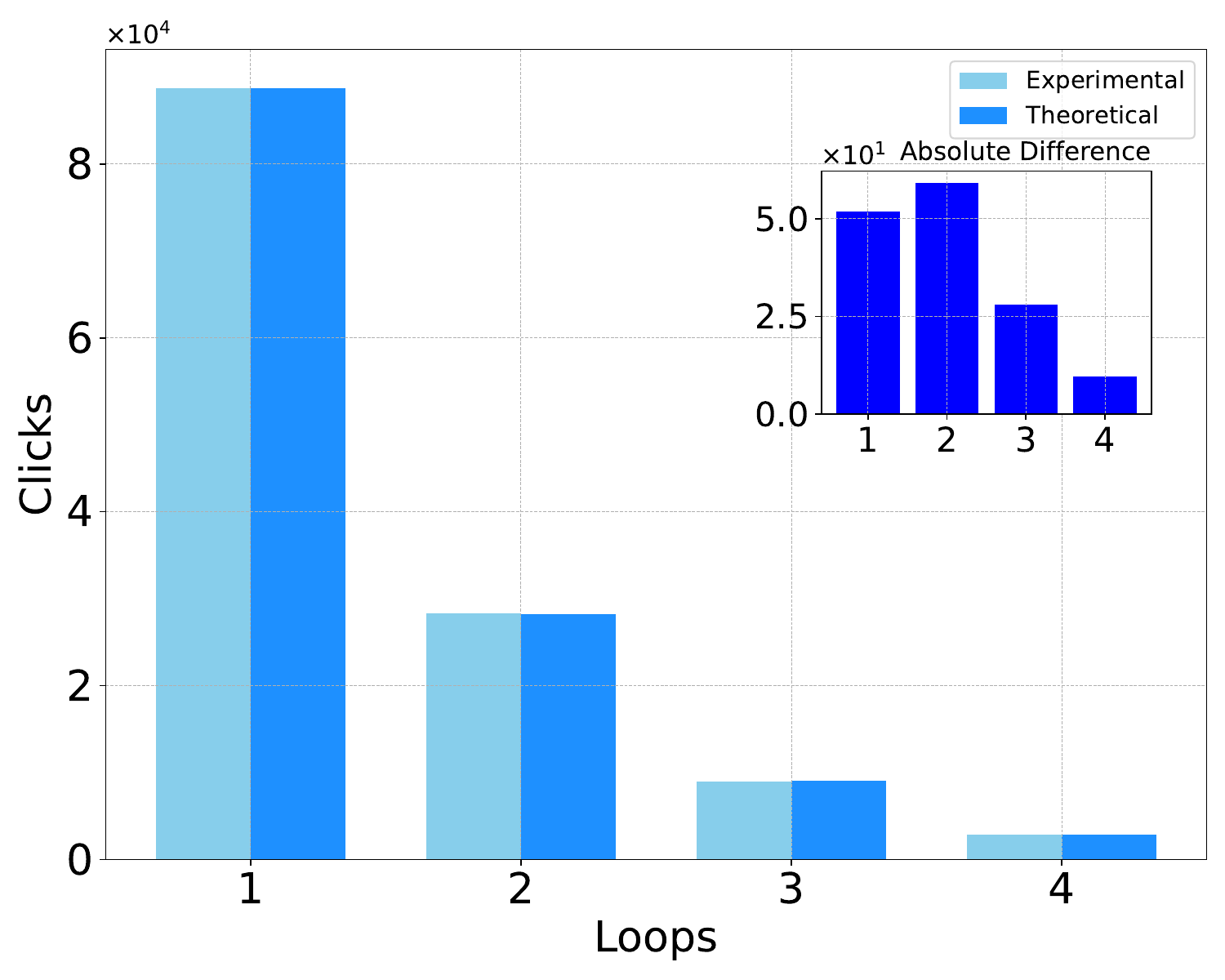}
    \caption{Comparison between experimental detection probabilities and theoretical predictions using the measured parameters $r=0.41$ and $\eta=0.230$. The close agreement confirms that the experimental data follow the model described by Eq.~(6).}
    \label{fig:comparision_t_e}
\end{figure}

Data were acquired in $0.6$-second intervals, resulting in a detection rate of approximately $1.2\times10^5$ single-photon clicks per second and a total dataset of about $17.41 \, \text{Mbits}$. The beam splitter reflectivity $r$ and loop loss $\eta$ were measured experimentally and inserted as input parameters into the theoretical model. The measured values, $r = 0.41$ and $\eta = 0.230$, were then used to predict the four detection probabilities $P_l$. The experimental detection probabilities, as shown in Fig.~\ref{fig:comparision_t_e}, follow the theoretical predictions closely, confirming that the observed statistics are consistent with the expected exponential dependence $p^l_{\text{click}} = 1 - e^{-\beta_l}$ described in Section~II.

\begin{figure}[ht]
    \centering
    \includegraphics[width=8.3cm]{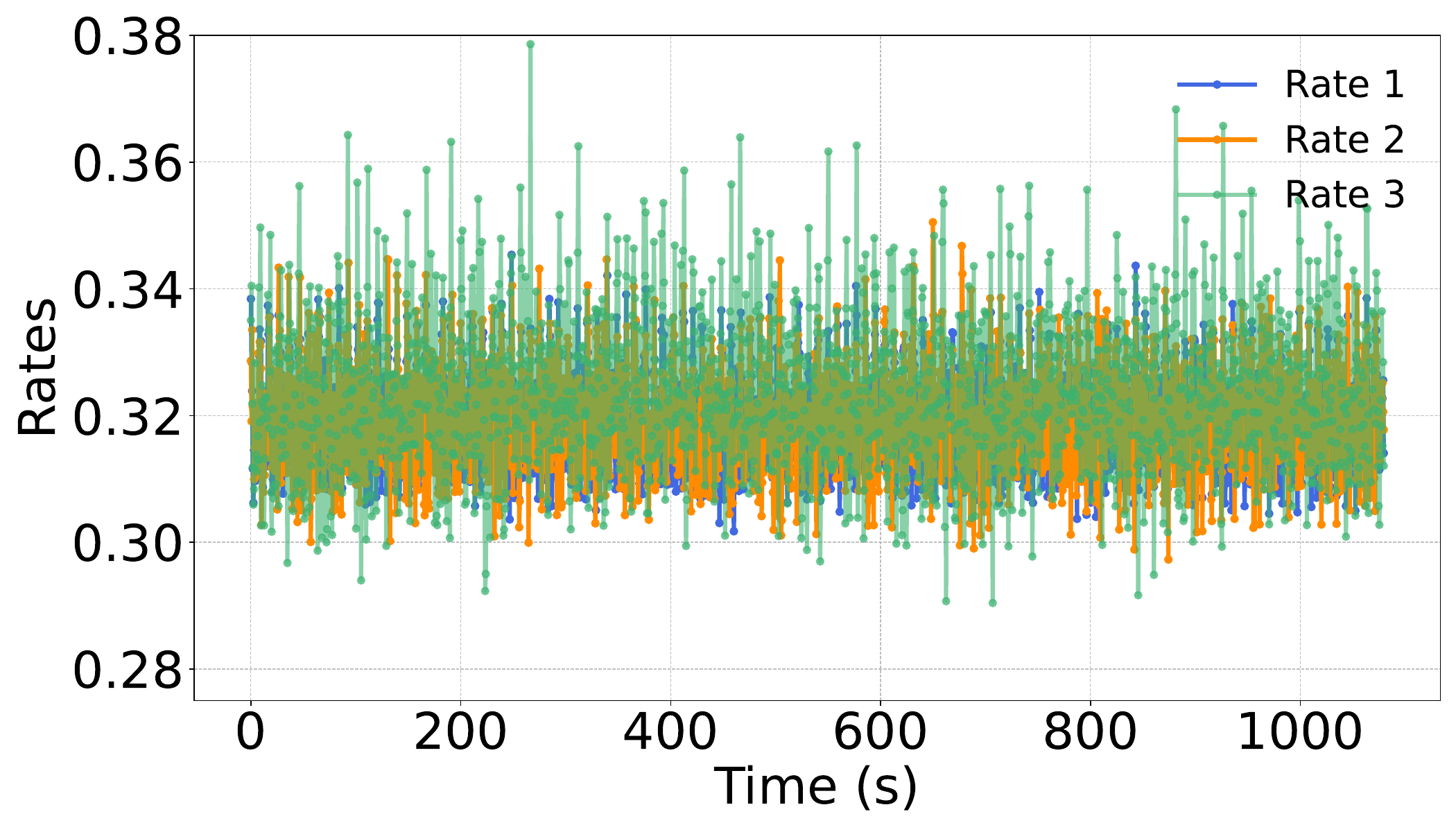}
    \caption{Detection rate ratios across different loop transits calculated over $0.6$-second intervals. The stability of the ratios confirms the constant relation $P_{l+1}/P_l \approx r(1-\eta)$ predicted by the model. Minor fluctuations in the third loop are attributed to lower detection statistics.}
    \label{fig:rate}
\end{figure}

To verify the temporal stability of the system, we analyzed the relative detection rates for each loop over consecutive $0.6$-second intervals. The ratio $P_{l+1}/P_l$ was found to remain constant throughout the measurement, in agreement with Eq.~\eqref{eq:ratio_sigle_det}. As shown in Fig.~\ref{fig:rate}, the measured loop ratios are stable over time, with small deviations at higher loop numbers attributable to statistical fluctuations at lower count rates.

Following the same procedure described in Section~II, the raw detection events were sorted into two disjoint bit sequences: $S_{Q_1}$ (private) containing events from $l \in [1,2]$ and $S_{Q_2}$ (public) containing events from $l \in [3,4]$. The resulting files contained respectively $15.8\,\text{MB}$ and $1.61\,\text{MB}$ of data, corresponding to $15.8$ and $1.61$ million one-bit samples. These data volumes exceed the minimum recommended sizes in SP~800-90B, which require at least one million bits for valid statistical testing, ensuring that both sequences are suitable for reliable evaluation.

% the difference arising from the lower probability of higher-loop detections.  

\begin{figure*}[t]
    \centering
    \includegraphics[width=15cm]{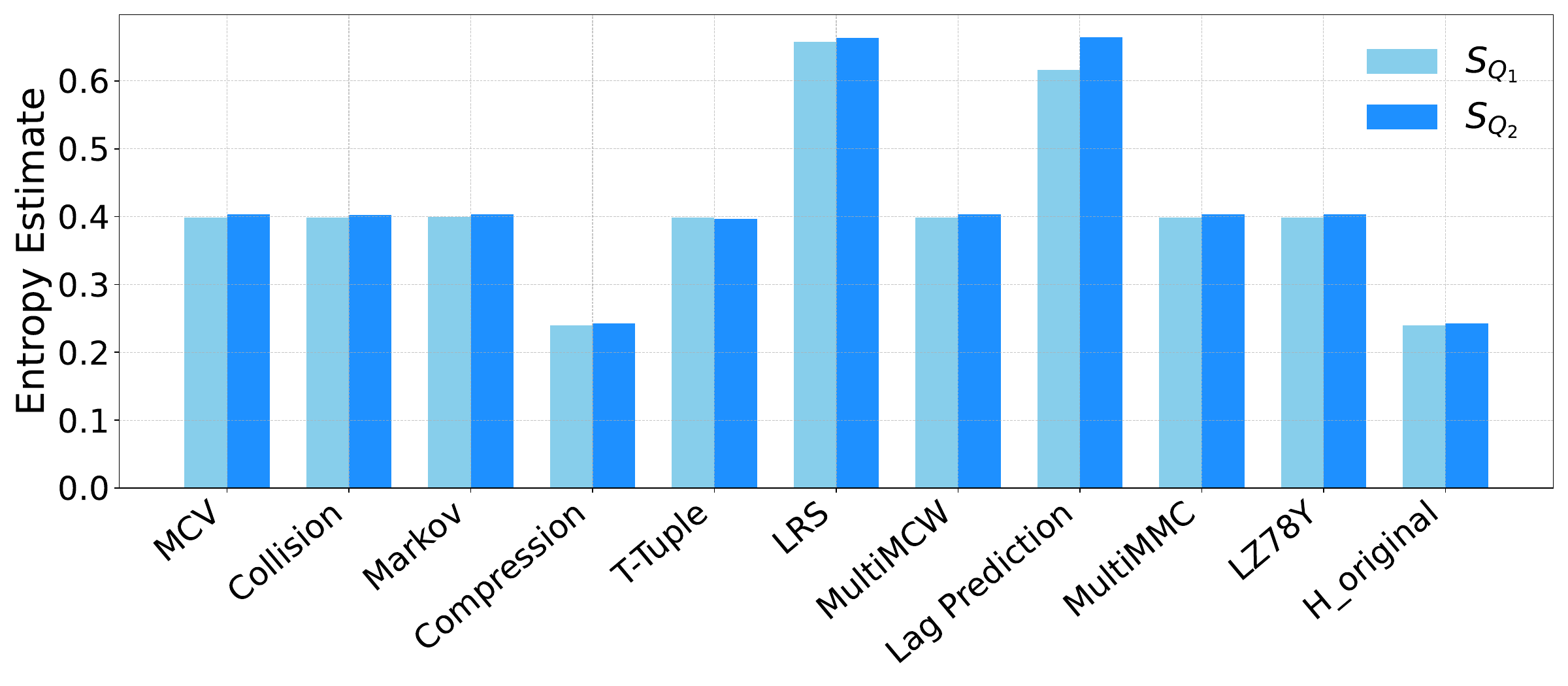}
    \caption{Results from the NIST test suite for non-iid sequences applied directly to the raw data of $S_{Q_1}$ and $S_{Q_2}$. The entropy estimators show close agreement, demonstrating that both sequences have similar statistical properties prior to post-processing.}
    \label{fig:nist_test}
\end{figure*}

In addition, both sequences were evaluated without post-processing using the official NIST entropy assessment tool for non-independent and non-identically distributed (non-iid) data. This tool estimates the worst-case min-entropy, and the results are expressed in bits of entropy per sample. The tests were applied directly to the raw data to compare the statistical properties of $S_{Q_1}$ and $S_{Q_2}$. As shown in Fig.~\ref{fig:nist_test}, all entropy estimators yield comparable values for both sequences, confirming that they possess nearly identical statistical behavior. This agreement validates the model’s prediction that the private and public outputs share the same entropy characteristics before randomness extraction.

\section{Discussion}

In this work, we addressed key aspects of robustness and verifiability in quantum random number generators through a design that combines a dual-sequence generation approach with an internal consistency check. The system emphasizes simplicity and stability, minimizing the number of active components and reducing potential points of failure. This simple architecture enhances reliability, a crucial requirement for cryptographic applications where long-term performance and reproducibility are essential for maintaining security.

The self-testing feature continuously monitors the system's internal statistics to ensure it behaves according to the expected quantum model. In our implementation, this corresponds to verifying that the measured detection probabilities satisfy Eq. \ref{eq:ratio_sigle_det}. This internal consistency condition allows the device to detect in real time any deviations caused by component degradation, environmental perturbations, or other non-ideal effects. When the condition fails, the system can flag or halt operation before compromised random numbers are produced.

In addition to self-testing, our design incorporates a public verification mechanism based on the generation of two separate sequences, $S_{Q_1}$ and $S_{Q_2}$, from distinct subsets of detection events. The two sequences are derived from disjoint outcomes: $S_{Q_1}$ from lower-order loop detections, and $S_{Q_2}$ from higher-order ones; hence, no event contributes to both sequences. Although their samples differ, both follow the same underlying probability distribution, yielding equivalent entropy properties. This symmetry allows one sequence to be safely released for public verification without revealing information about the private sequence. Experimentally, this equivalence was confirmed by the NIST non-iid entropy tests, which showed matching statistical properties between $S_{Q_1}$ and $S_{Q_2}$, validating the public-verification concept.

The two verification layers, self-testing and public validation, jointly strengthen the QRNG’s reliability and transparency. The self-testing mechanism provides an internal safeguard ensuring proper device operation, while public verification enables external users or auditors to confirm output quality without accessing private data. Together, these mechanisms demonstrate a practical path toward trustworthy and continuously verifiable quantum randomness generation.

Future implementations could explore modified geometries, for example, introducing an additional beam splitter before the detection port to prevent the first reflection from directly reaching the detector. Such a configuration would mitigate potential detector blinding and allow finer control over the effective reflectivity without altering the self-testing condition, further improving the operational robustness of the device. 

In this work, we did not explicitly model the contribution of detector dark counts or other noise sources. Nevertheless, if such effects were significant, they would manifest directly in the monitored detection ratios used for self-testing. In particular, an excess of dark counts would appear as an anomalous increase in the last detection rate, corresponding to the highest loop number, which is the most sensitive due to its lower photon count. This behavior could therefore be identified through the same consistency check employed for system monitoring, reinforcing the diagnostic capability of the method. A quantitative analysis of how different noise levels influence the rate ratios and the entropy of the sequences will be an important direction for future investigation.

\bibliographystyle{ieeetr} 
\bibliography{ref.bib}

\end{document}